# Photonic hook - a new type of subwavelength self-bending structured light beams: a tutorial review

Igor V. Minin[1,2], Oleg V. Minin[1,2], Liyang Yue[3], Zengbo Wang[3], Valentyn S. Volkov[4,5], and Demetrios N. Christodoulides[6]

[1]Tomsk Politechnical University, Tomsk 634050, Russia
[2]Tomsk State University, Tomsk 634050, Russia
[3]School of Electronic Engineering, Bangor University, Dean Street, Bangor, Gwynedd, LL57 1UT, UK
[4]Moscow Institute of Physics and Technology, Institutskyi 9, 141700, Dolgoprudnyi, Russia
[5]SDU Nano Optics, University of Southern Denmark, Campusvej 55, DK-5230, Odense, Denmark
[6]College of Optics/CREOL, University of Central Florida, Orlando, Florida 32816, USA

**Abstract**
During the last 2 years, it was shown that an electromagnetic beam configuration can be bent after propagation through an asymmetrically shaped (Janus) dielectric particle, which adds a new degree of simplicity for generation of a curved light beam. This effect is termed "photonic hook" (PH) and differs from Airy-family beams. PH features the smallest curvature radius of electromagnetic waves ever reported which is about 2 times smaller than the wavelength of the electromagnetic wave. The nature of a photonic hook is a the dispersion of the phase velocity of the waves inside a trapezoid or composed particle, resulting in an interference afterwards.

Compiled: April 18, 2019

## 1. Family of Airy beams

The idea that light propagates along straight lines is known since the days of ancient Greek philosophers. The development of Maxwell's electrodynamics further reinforced these notions by ensuring the conservation of electromagnetic momentum. The possibility that a wave packet can freely accelerate even in the absence of an external force was first discussed four decades ago within the context of quantum mechanics [1]. As indicated in [1], this is only possible as long as the quantum wave function follows an Airy-function profile. In 2007, this Airy self-acceleration process was first suggested and experimentally observed for the first time in optics [2, 3]. It is important to note that the Ehrenfest theorem still holds thus preserving the balance of the transverse electromagnetic momentum for finite power Airy beams while the local intensity features do self-bend in a self-similar fashion (Figure 1). Ever since, this class of accelerating or self-bending beams has attracted considerable attention and found applications in many and diverse fields, especially in the paraxial domain. In the past few years, other types of Airy-like accelerating curved beams have been intensely explored; among them: "half Bessel" [4], Weber (travel along parabolic) and Mathieu (travel along elliptic curves) beams [5,6]. These beams are by nature non-paraxial and hence can bend at larger angles. In all cases, these Airy-like wavefronts propagate on a ballistic trajectory over many Rayleigh lengths while defying diffraction effects. Until recently, they provided the only example of "curved light transport" in nature.

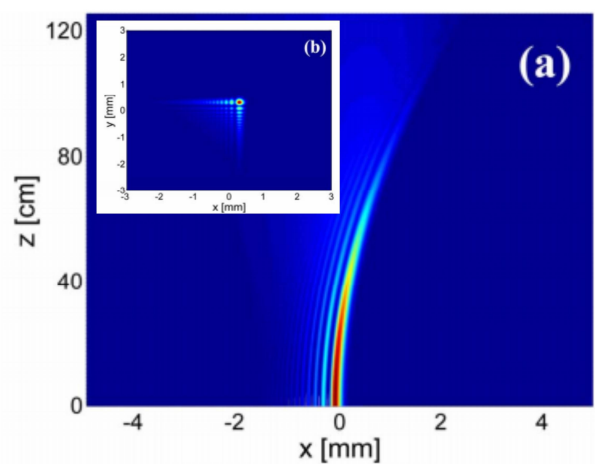

Fig.1. (a) Propagation dynamics of a finite energy typical Airy beam as a function of distance (b) Two-dimensional finite energy Airy beam after propagating z =50 cm. Adapted from [14]

A common optical schematic of the Airy beam formation includes a cylindrical lens and a spatial light modulator (SLM) which operate only at restricted power. The latter is mounted in the front focal plane of the cylindrical lens, while the Airy beam generates in its back focal plane and nearby. Cubic phase modulation of an incident Gaussian beam generated from modulator acts as a cubic lens, whereas the curved path is determined by the acceleration rate of the cubic phase change [2]. Airy laser beams are generated by diffractive optical elements in transmission [7] and reflection [8] modes. Diffraction gratings are used as SLM analogs [9, 10], which create bending diffraction orders. Metasurfaces [11] are also used for generation of Airy beams, however, they are rather complicated and expensive.

It is important that the Airy beam's characteristic diameter is related to the lens numerical aperture, usually amounting to tens of wavelengths, with the Airy beam's path length being related to the optical element diameter [2,3].

At the same time, Airy beams usually generate by a composite optical element with a cubic phase or SLM positioned behind the focus of spherical lens. It is worth noting that scaling the Airy beam generation from optical to, for example, terahertz range, is not always possible. This is because SLMs do not operate in the terahertz range due to the absence of materials with required modulation [12]. Moreover, the main lobe of a finite energy Airy beam is not observable directly behind the cubic phase element and a transition region exists, where the initial intensity distribution of the incoming beam is transformed into the distinct Airy pattern [13]. The modern state-of-art of Airy beams developments were recently reviewed in [14].

From the geometrical optics point of view, for conventional Airy-family beams, their paths are formed by the caustic envelope of a family of tangential geometrical rays, indicating that the maximum bending angle is determined by the steepest ray and, thus have a small curvature [15].

## 2. Photonic jet

As it well known, the interaction of light with transparent spherical particles has been heavily investigated over the years since the days of Pliny the Elder (AD 23–AD 79) [16]. In 2004, Chen et al. coined a new term "photonic nanojet" (PJ) for the sub-wavelength-scale near-field non-resonant light localization at the shadow side of a mesoscale (with dimensions of several units of wavelength) low-loss dielectric cylindrical or spherical particle [17]. By increasing the dimensions of a spherical particle the electromagnetic field structure evolves and tends to be more localized and directed forward. Later it was shown that this phenomenon can also be observed for three-dimensional particles of arbitrary shape even with about one-wavelength diameter [18]. However, despite the apparent simplicity of the problem, the physics of the formation of a photonic jet by mesoscale particle is quite deep [19]. Importantly, mesoscale dimensions of a particle involves the interaction of light with dielectric structures of intermediate scale - structures that are too small to be described by traditional continuum methods and too large to be characterized as simple dipoles [20]. Moreover, despite the beneficial performance of PJs in several applications, up to date, all photonic nanojet beams are fundamentally similar - single-colored and straight. It is important to noted that due to PJ could maintain a subwavelength focusing jet along the path length more than two wavelength beyond the shadow surface of dielectric particle there are a wide-spread but misleading claim that PNJ is a non-evanescent propagating beam within which evanescent wave doesn't contribute [17]. In fact, evanescent waves could play strong role in near-field PNJ.

## 3. Photonic hook

Recently, a new family of near-field localized curved light beams has been discovered by Minin and Minin, which differs from the family of Airy beams and is formed in the near-field zone: the so-called "photonic hook" (PH) [20].

In a photonic hook, individual light waves propagate along a straight line, but their relative phases and amplitudes are chosen so that the maximum energy in space forms a constant curved shape in time. Since the electromagnetic waves themselves propagate along linear paths, the energy density in the curved beam decreases with distance from the source. It could be noted that the PH phenomenon is non-resonant in nature but scattering wave in this area is not only transverse but contains both transverse and longitudinal components.

Our recent research shows [21] that it is possible to produce a localized optical beam whose intensity peaks move along curved trajectories in the near-field (Fig.2a). Unlike ordinary optical localized beams (like PJ), the photonic hook self-bends (i.e., transversely accelerates) throughout propagation in the near-field. PHs appear even in free space and do not require any waveguiding structures or external potentials. They are unique in the sense that their radius of curvature is substantially smaller than the wavelength, meaning that such structured light beams have the maximum acceleration among the known curvilinear beams. The subwavelength curvature of electromagnetic waves with such a small radius was described for the first time in Refs.[20,21]. It is shown (Fig.2b) that that minimal beamwaist of the photonic hook is shorter than that for the photonic jet induced by the symmetric particle, and is able to break the simplified diffraction limit (0.5λ criterion) [21]. It could be noted that in the nonparaxial approximation Bessel like beams with their conical-interference structure may follow hyperbolic or hyperbolic secant trajectory with large bending angles and subwavelength FWHM of main lobe. Such characteristics of Bessel-like beams were observed when the phase at the input plane is engineered so that the interfering ray cones are made to focus along the prespecified path [22]. But the curvature of such a beams is more than several wavelength.

The photonic hook is formed in the spatial region where the effects of near-field scattering play a significant role. Typically, this near-field region is located at the distances not exceeding 1-2 particle dimensions and is characterized by marked contribution of the radial component of optical field. In turn, this condition imposes limitations on the range of dielectric particles sizes, so it must be about a few wavelengths and even equal to the radiation wavelength, i.e., have a mesoscale dimensions [20].

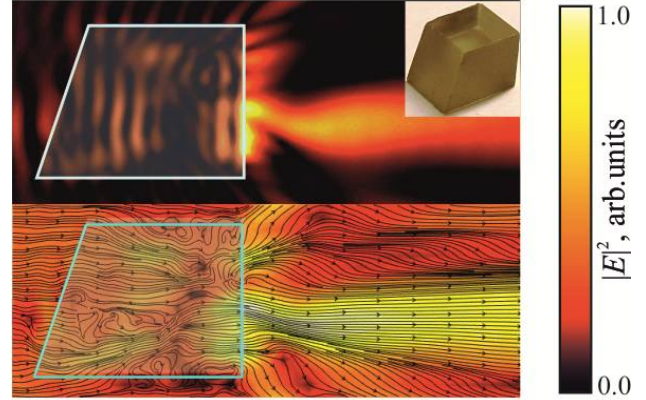

*Fig.2a. Simulations of a photonic hook formation (Field intensity and Power flow)*

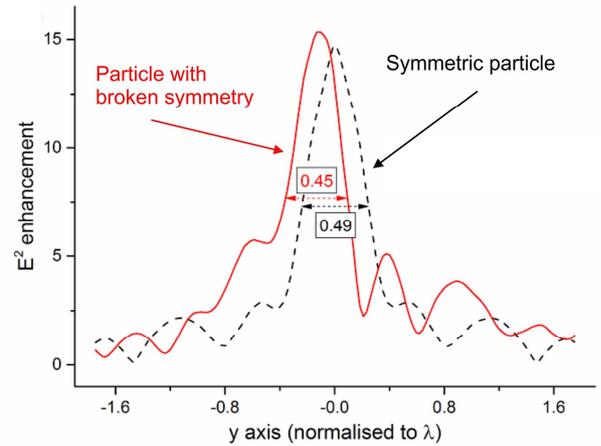

*Fig.2b. $E^2$ enhancement profiles along the y axis for the symmetric and asymmetric cuboid particles with L = 2.5λ at the plain of minimal beamwaist.*

The realization of such curved near-field beams is possible based on the interaction of a plane wave with a mesoscale dielectric particle with broken symmetry.

In [20, 23-28] we theoretically studied a PJ formed behind the symmetric cubical particle with an equal rib and discussed some examples of PHs behind the mesoscale particles with broken symmetry [20]. The curve of the brightest region results from a complex near-field interference pattern inside the particle as the phase velocity disperses. Because of this shape, the time of the full oscillation phase of an optical wave varies in the particle unevenly. As a result, a curved light beam is produced at the exit from the particle (near its shadow surface) [20,21].

In the other words, most of the power flow distributions of the normally straight photonic jets only match the profile of the high-intensity area in the region where power flows just exit the particle [29]. However, as it was shown in [21] the photonic hook fully match the trend of power flows in a certain shadow region. The vortices in the power flow are stable foci in the phase space [20], and the power flow couples to the other planes through the singular points at the center of the vortex [31]. Interference caused by the diversity of the particle thickness along the polarization direction produces some more singular points in the vicinity of the asymmetric particle, which induces the curvatures appearing at the streamlines.

Besides, the curvature of the beam can be adjusted by changing the wavelength, polarization of the incident light, as well as the geometric parameters of the dielectric particle [21]. Such flexibility is likely to widen the future applications of this approach.

The photonic hook are highly asymmetric localized near-field and as a result their energy is more tightly confined in one quadrant thus increasing the energy density in the main lobes.

One of an idiom supported by people since ancient times is ''Seeing is believing''. So the experimental verification of this accelerating near-field beams would be an additional proof of the PH concept, which can be used to control light at the new level.
On the other hand direct experimental confirmation of this phenomenon is not trivial due to both difficulties in fabrication of micro-scale particles with high precision and the limited resolution of existing 3D sub-wavelength-resolution imaging modalities in the visible range. Moreover, in a realistic optical band system, for handling the asymmetric dielectric particle, it would need to be in contact with a finite-size dielectric substrate, which generates its own scattered field, interacting with that of a particle, thus leading to image distortions [32].

To this end, recently, the existence of Minins' PH phenomenon has been verified experimentally for the first time [33] using continuous-wave scanning-probe microscopy, operating at 0.25 THz due to the scalability of Maxwell's equations. As a material platform for fabrication of dielectric particle we select polymethylpentene (TPX, the refractive index is n=1.46 at 0.25THz). It was found that the TPX cuboid featuring the rib length of L=4.8mm and the prism angle of θ=18.4º ensures generation of PH with the maximal curvature.

In Fig. 3, we show the results of experimental PH visualization [33]. It can be seen that curvature of PH is about $\alpha =148^0$, the length of PH is less than two wavelengths with inflection point position near Z=1.2λ from the shadow surface of the particle. In contrast to traditional Airy beams (generated using a complicated optical element with cubic phase or with a spatial light modulator behind the focus of a spherical lens [2,3]), a PH can be created using a compact mesoscale dielectric particle-lens. Moreover, the Airy beam consists of the main lobe and a family of side beamlets whose intensity decay exponentially [2,3]. Interestingly, in the case of PH, only the main lobe has a curved shape, and the family of curved sidelobes is absent. It was also shown that PH phenomenon is observed on a scale much smaller than Airy beams [21,33,34].

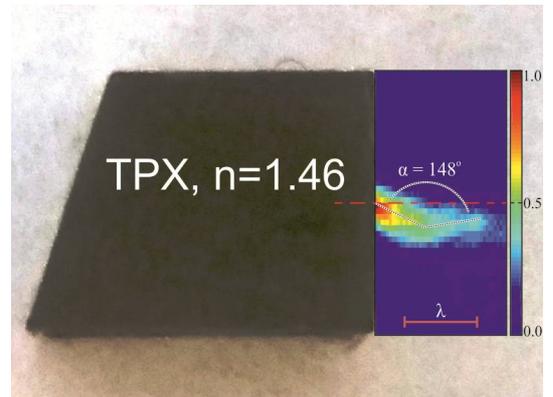

*Figure 3. Experimental visualization of a photonic hook: a photo of dielectric particle and THz field intensity, formed at its shadow side and illustrating PH effect.*

It was also experimentally confirmed [33] that the lateral dimensions of photonic hook are also subwavelength with FWHM of main lobe less than diffraction limit.

## 4. Janus particle and photonic hook

Janus particles are named after the Romanian mythology god Janus. His two opposing faces peer to the past and the future, to creation and destruction. The term of "Janus particle" have received wide attention since they were discussed in de Gennes' Nobel lecture in 1991 [35] and the terminology is based on the special architectural feature of having two sides or at least two surfaces of different physical or/and chemical and/or polarity.

These particles are unique among micro objects because they provide asymmetry and can thus impart drastically different physical properties and directionality even with in a single particle. The broken symmetry offers efficient and distinctive means to realize the emergence of properties in unconceivable for homogeneous particles or symmetric particles.

From this point of view, since the formation of a photonic hook occurs on the basis of a particle with broken symmetry, such a particle may be considered as a Janus particle. The curved electromagnetic beam caustic is formed only when the asymmetric particle is irradiated in one direction - if the particle is pumped from the side of the prism (and not in the opposite) due to broken symmetry [20]. Also the shape of the dielectric particle considered above is not the only possible one.

Indeed, direction of illumination appears to be a crucial factor determining the PH existence. It was surprise that a similar effect to that observed under the illumination of the configuration shown in Figure 2a, at 90 degrees (see Figure 4a below).

From the simulation results it clearly follows that the structure of the photonic hook is destroyed when the particle is inclined. However, when radiation is incident from the side of a particle with a shorter edge length, a photon hook is formed at the exit from the particle near the edge with a greater length. When irradiated in the opposite direction, a photonic hook is not formed. Also for all other orientations of illuminations, the PH strongly distorts or even disappears.

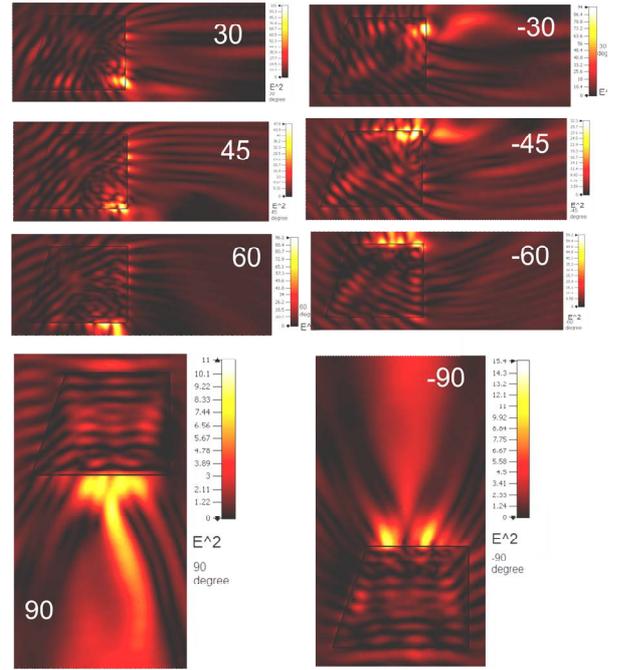

**Figure 4a.** *Formation of curved focusing beams under the oblique illumination of configuration from Fig.1.*

Therefore, these different effects of the trapezoid particle illuminated by the waves from the multiple angles, as shown in Fig. 4a, demonstrates a common rule for photonic hook generation, which can be summarized as follows: photonic hook is formed during the incident wave propagates in a trapezoid particle, from its narrow end to the wide end, because disperse of phase velocity for the wave can be maximumly enhanced through enlargement of cross-section of the high-index dielectric material.

A new way to generate PH using Janus particles is to violate the symmetry of the material inside the microparticle. By changing the refractive index contrasts between adjacent materials of the composite particles or/and rotating the Janus microparticle as a whole relative to the direction of radiation incidence, the PH shape profiles can be effectively controlled, as shown below in Fig.4b.

PH also has a cross-dimension smaller than the wavelength that makes high resolution possible. Thereby, the observed PH phenomenon can provide advantages over common photonic jets in imaging applications, similar to the beneficial character of the Airy beams over the Gaussian and Bessel beams reported in Ref. [36].

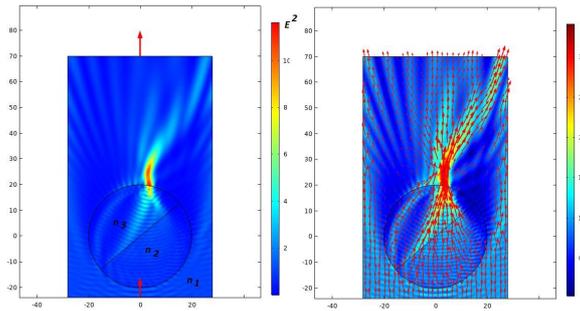

***Figure 4b.*** *Formation of curved focusing beams under the oblique illumination of composed cylindrical particle from two materials with different refractive index: left- field intensity distributions, right – Poyting vector flow.*

## 5. Optimechanical manipulator and optical switch

Taking into account the properties of a photonic hook mentioned above, it turned out that the photonic hook based optomechanical manipulator can guide the nanoparticle in a curved trajectory even around glass obstacles [37]. The changes in the photonic hook's optical forces when relatively large glass and gold obstacles were introduced at the region where the curved photonic jet is created were also considered. It was show, that despite the obstacles, perturbing the field distribution, a particle can move around glass obstacles of a certain thickness. For larger glass slabs, the particle will be trapped stably near it. The revealed new optomechanical effect [37] allows moving particles around a specific path [38], paving the way to enhanced and more flexible optical manipulation of nanoparticles and their transport along non-straight trajectories without the complicated employment of classical Airy beams.

An optical switch is a device that selectively switches optical signals on or off or from one channel to another. The 'photonic hook' effect may be used for nanoscale light switching and guiding, for example, in a Photonic Integrated Chip. As it was shown in [21], a dielectric trapezoid particle illuminated by a plane wave can form a bent light beam and its curvature is proportional to ratio between particle size and incident light wavelength in a certain range. According to the above principle, a size-fixed dielectric trapezoid particle combined with cuboid and prism should function like an optical swtich for light-route selection depending on incident light wavelength. It is proved that the incident light with a wavelength that is 33% of particle length (short wavelength light in ***Fig.5***) significantly deviates its original straight propagation route and creates a photonic hook with a 35° deflection angle after passing the particle. Otherwise, the same trapezoid particle cannot bend the light with a wavelength as long as its length (long wavelength light in ***Fig.5***), and the incident light maintains its straight propagation route as a photonic jet. Based on this phenomenon, a dielectric trapezoid particle can simultaneously guide two lights varing wavelength to the different light route in the form of photonic jet or photonic hook for the purpose of frequency (wavelength) switching, as shown in ***Fig.5***.

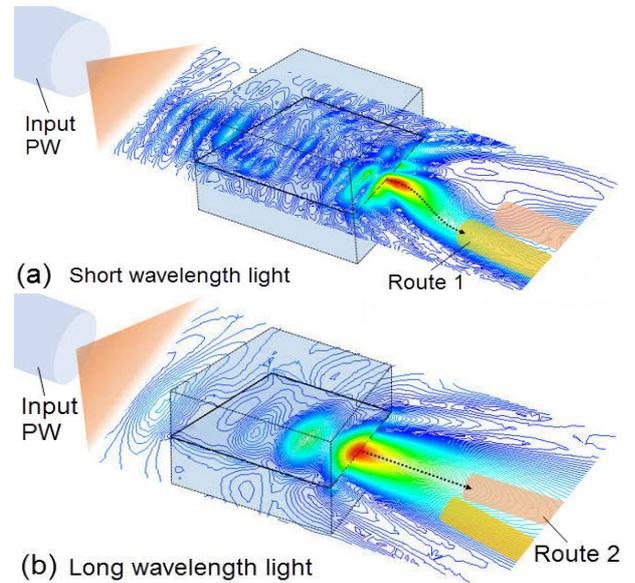

Fig.5 Diagram of a particle optical-switch (a) short wavelength light – 33% of particle length (b) long wavelength light – 100% of particle length.

## 6. Caustic-based curved localized near-field beams

Simulation and experimental results of light with wavelength of 600 nm and 400 nm diffracted across a corner of 1/3 and ½ of wavelength heights thick film respectively (with refractive index n= 1.5 for both film and substrate) were reported firstly in [39]. It was shown the diffraction angle and width for the incident 600-nm wavelength (corner height of 1/3 of wavelength) were larger than those for the incident 400-nm wavelength (corner height of 1/2 of wavelength). It has been also shown that light are split and diffracted substantially at the edge, even when film thickness was relative small and the field localization area near the corner has a curved shape. Light diffracted more to the higher-index region (film) than to the lower-index region (air).

On the other hand it is well known that under diffraction on dielectric edge in the Fresnel approximation the edge wave has the eikonal S ~ Z+X$^2$/2Z) [40] and a solution of the diffraction at the edge of the semi-infinite opaque screen [41] is a paraxial 2D light for which the argument of the complex amplitude function is depends on variables, like X$^2$/Z, where X is the transverse coordinate, and Z is the longitudinal coordinate.

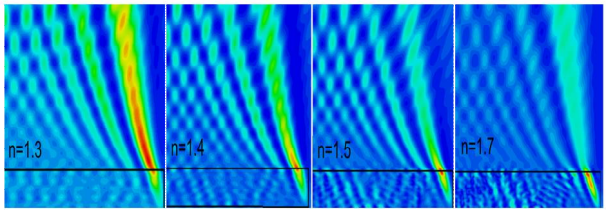

Fig.6. Edge diffraction on phase plate.

It means that instead of the Airy beams which have a parabolic path y=x$^2$, whereas the considered edge beams in question propagate along a root-parabolic path y=sqrt(x).

Pioneering publications on the formation of the photonic hook [20,21,38] have initiated research into alternative approaches. For example, the curved PJ based on off-axis illumination of spherical particle was studied in [42].

Light passes through the upper part of the microsphere and forms a curved envelope of light rays near the surface of microsphere due to the large spherical aberration of the microsphere causes significant caustics [43].

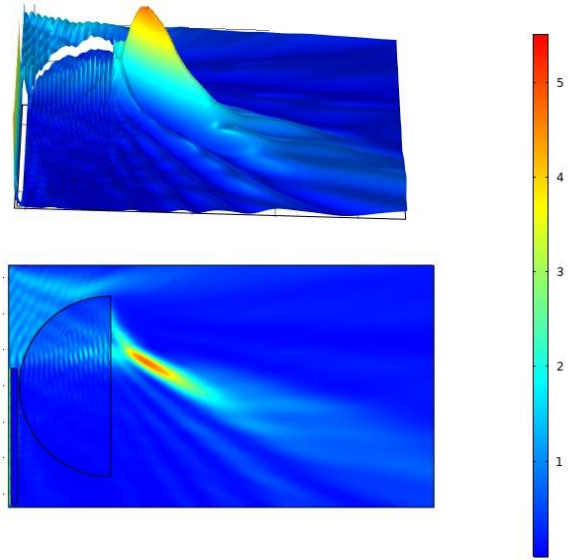

Fig. 7. Formation of Airy-like beam by hemispherical particle.

Bernhard Baruch in *New York Post* (24 June 1965) say: "*Millions saw the apple fall but Newton was the one who asked why*". However, no one argued that the caustic curve in the case described above is an Airy-like beam formed near the shadow surface of dielectric cylinder. Fig. 7 show formation of an Airy-like near field beam under off-axis illumination of hemispherical particle.

Another way to get the curved photonic jets was considered in [44]. A glass cuboid was embedded inside a dielectric cylinder when illuminated with a monochromatic plane wave. As a results near the shadow surface of cylinder the curved PJ was observed. This can be understood as an interference phenomenon, which takes place in the caustic and is caused by the break of the symmetry. Another interpretation corresponds to an analogy with off-axis aberration, mainly coma.

The effect of curved photonic jet later was also observed for four-layer microcylinders. Due to for relatively small angles between the incident wave and the front surface of the multilayered cylinder, streams cannot enter the core region of cylinder. They converge with the upper and bottom substreams from the core region leading to two high-intensity photonic curved jets formed beside the ultra-narrow PJ. The formation of the photonic nanojet and two curved PJ could be regarded as an analogue of optical super-oscillation, where a super-narrow hot spot with high-power side lobes around it is formed [45].

Nevertheless, the acceleration (curvature) of photonic jets on the basis of the approaches considered above is much less than on the basis of a cuboid with broken symmetry. Moreover, the subdiffraction value of the main beam lobe was not observed yet.

### 7. Photonic hook plasmon

The concept of photonic hook has gone beyond optics and penetrated other fields, including plasmonics. In 2010, the idea of Airy surface plasmon polaritons (SPPs) was introduced [46], and experimental observations soon followed [47].

It could be noted that owing to the limited propagation length of surface-plasmons, the resulting beams should be formed directly in the near-field, before they decay. Moreover, unlike planar phase plates, surface-plasmons are excited over a finite propagation distance and therefore their phase cannot be simply defined at a specific one-dimensional plane.

Different methods for Airy SPP excitation were demonstrated: by directly coupling 1D Airy beams from free space to SPPs on a metal surface through a grating coupler [48], by the use of a grating pattern designed to imprint the phase profile of an Airy function [49], by SPP incidents into a nonperiodically arranged nanocave array [50], produced by a wedged metal–dielectric–metal structure [51], by a metasurface, based on the amplitude and phase modulation of subwavelength slits [52], by SPP propagations along the interface between a dielectric and a negative-index metamaterial [53], generation of Airy plasmons on a graphene surface [54], all-fiber Airy-like beam generator based on a 1D groove array on the gold deposited end facet of an optical fiber [55], based on paired nanoslit resonators on a metal surface [56], etc. Also, plasmonic Airy and Weber beam (which is a rigorous non-diffracting solution of the 1D Helmholtz equation in parabolic cylindrical coordinates) based on curved nanoslit array, was studied in [57]. In all cases, permanent and difficult-to-fabricate nanostructures are required.

However, in low-dimensional systems (in which at least in one of the three dimensions of the electronic state wave function is confined) until recently the families of Airy plasmon beams were the only beams that have a curved trajectory. But for Airy-type SPPs high beam acceleration is required to achieve significant curvature of the beam over the propagation length.

Due to the vectorial problem of surface plasmon propagation at a metal-dielectric in-plane interface may be formally reduced to a scalar one, plasmonic hook recently was first offered in [58]. This concept is fundamentally simpler than the generation of the SPP Airy-family beams.

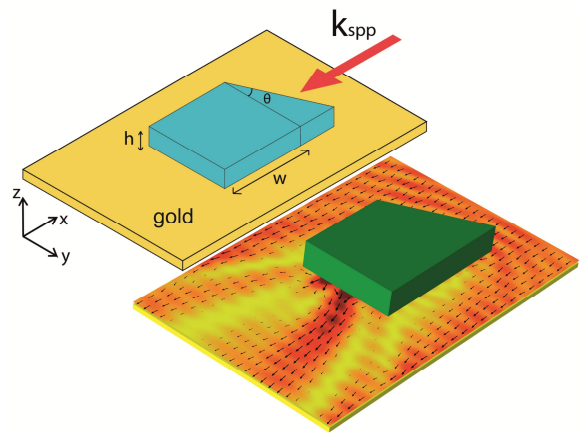

Fig.8. Photonic hook plasmon formation.

As in the case of free-space photonic hooks, plasmonic hooks represent the only plasmonic wave that is self-bending at subwavelength scale - the PHP propagates along wavelength scaled curved trajectory with a radius less than the SPP wavelength, which represents the smallest radius of curvature ever recorded for SPP beams, and can exist despite the strong energy dissipation at the metal surface. In addition, it has been shown that plasmonic hook can be excited over a broad set of wavelengths and accelerating of a plasmonic hooks (curvature) depends on the illuminating wavelength [58]. It was also demonstrated that the plasmonic hook can be guided along a metallic interface without using any physical waveguide [59] structures.

Importantly, it also demonstrated for the first time a simple and reliable real-time technique to dynamically control the curved trajectories of light beams over metallic surfaces by simply changing the illumination wavelength (without the need of any permanent guiding structures) [58].

It is worth to note that the plasmonic PH is formed in the spatial region where the effects of evanescent fields play a significant role. Because of their wave nature, SPPs also undergo diffraction in the plane of the interface. The use of overlapping evanescent fields from several active radiators causes the specific interference interaction between the radiators which allow breaking the fundamental Rayleigh criterion.

From an applications perspective, photonic hook plasmons are important for the following main reasons. First, it demonstrates that the Airy plasmons now are not the only self-bending beams in two dimensions. Second, the photonic hook plasmons are self-bending localized field, giving us the means to control the flow of light on the surface in the near-field at subwavelength scale. Moreover, electromagnetic waves at the surface of a metal can be channeled into circuit components smaller than the diffraction limit.

## 8. Future principal possibilities

Because of the phenomenon of focus bending the dielectric particle is caused by the interference of waves inside it as the phase velocity disperses the shape of the photonic hook and the characteristics of curved near-field beams, depending on the specific application, can be quite exotic due to the control of phase delays across the wavefront by choosing the shape of mesoscale particles (Fig.9).

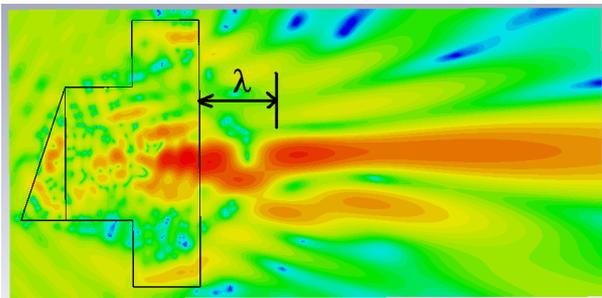

*Figure 9. Example of a photonic hook in the form of a loop.*

## 9. Acoustic hook

It is well known that acoustic waves are elastic and longitudinal in nature waves. However, electromagnetic (optical) waves are transverse due to oscillating electric and magnetic fields perpendicular mutually and also to the direction of motion.

The observation of a new type of near-field curved acoustic beam different from the Airy-family beams both through simulations and experiment recently was reported [60]. This new self-bending acoustical beam also was formed from a rectangular trapezoid of a dielectric material immersed in water. The origin of this curved beam is in the vortices of intensity flow that appear inside the solid due to the conversion of the incident longitudinal wave mode to a shear wave in a solid. These vortices redirect the intensity flow resulting in a bending of the beam.

## 10. Summary

Importantly, these new possibilities prove that the concept of wave acceleration appears in general electromagnetism and can be observed in non-paraxial approximation in the near-field. Another important outcome is that the localized near-field beams no longer have to propagate only in a straight line - they can be self-bending [34].

Finally, there are a few reasons that the concept of photonic hook has gone above and beyond optics, penetrating other fields of modern physics (i.g., terahertz, plasmonics, and acoustics). First, the PH concept is universal - this phenomenon occurs under non-paraxial conditions in near-fields, thus opening interesting opportunities in other physical systems, i.g., not only electromagnetic, but mechanical wave too - acoustics and ultrasonics. Second, the early experiments have revealed that PH beams are exceptionally easy to realize and manipulate. Note that the mesoscale dimensions and simplicity of the PH concept open a way for the practical integration of PH elements into lab-on-a-chip platforms and indicating their large scale potential applications.

We believe that the curved trajectory of PH is beneficial to photonics and the related disciplines. It could be used for advanced manipulation of nanoparticles, biological systems, material processing, and surgical systems, where curved sub-wavelength-scale optical beams are of great interest. The method could accommodate other applications, including particle acceleration and separation, short-range micromanipulation, terahertz generation, near-field spectroscopy, laser machining of various guiding structures to include wavelength scaled division multiplexers, and interferometer beam-splitting and beam-coupling, to redirect an optical signal in meso- and nano-scale or even as a scalpel tips for potential use in ultraprecise laser surgery [61].

On a more fundamental level, the introduction of the photonic hook concept into the field of optics will help the scientific community to better understand the physical phenomena of accelerating near-field waves and to manipulate light in unusual ways [62]. As it was mentioned in [34] "such elegant examples of structured light beams deepen our understanding of light propagation, and will fuel anticipation and excitement for the next generation of imaging and manipulation". Moreover, it has opened up new paths for new applications.